%% file: 19-sc.tex
\documentclass[sigconf]{acmart}
\def\BibTeX{{\rm B\kern-.05em{\sc i\kern-.025em b}\kern-.08emT\kern-.1667em\lower.7ex\hbox{E}\kern-.125emX}}

\usepackage[section]{placeins}
\usepackage[outdir=./png/]{epstopdf}
\usepackage{times}
\usepackage{graphicx}
\usepackage{amssymb}
\usepackage{amsmath}
\usepackage{subfig}
\usepackage{epsfig}
\usepackage{psfrag}
\usepackage{multicol}
\usepackage{multirow}
\usepackage{amsmath}
\usepackage{mathtools, cuted}
\usepackage{algorithm}
\usepackage{algpseudocode}
\usepackage{flushend}
\usepackage{amsthm}
\usepackage{epstopdf}
\usepackage{caption}
\usepackage{xcolor}
\usepackage{amsopn}

\usepackage[normalem]{ulem}

\newcommand{\eat}[1]{}

\algrenewcommand\textproc{}
\algnewcommand\And{\textbf{and}}

\setcopyright{none}
\renewcommand\footnotetextcopyrightpermission[1]{}
 
\begin{document}

\title{uPredict: A User-Level Profiler-Based Predictive Framework \\ for Single VM Applications in Multi-Tenant Clouds}

\author{Hamidreza Moradi, 
Wei Wang, 
Amanda Fernandez and
Dakai Zhu}
\affiliation{
\institution{The University of Texas at San Antonio} 
\city{San Antonio}
\state{Texas}
\country{USA}
}
\email{ {hamidreza.moradi, wei.wang, amanda.fernandez, dakai.zhu}@utsa.edu}


\pagestyle{plain}
 
\begin{abstract}
  
  Most existing studies on performance prediction for virtual machines
  (VMs) in multi-tenant clouds are at system level and generally
  require access to performance counters in hypervisors. In this work,
  we propose {\bf \em uPredict}, {\em a user-level profiler-based
    performance predictive framework} for single-VM applications in
  multi-tenant clouds. Here, three micro-benchmarks are specially
  devised to assess the contention of CPUs, memory and disks in a VM,
  respectively. Based on measured performance of an application
  and micro-benchmarks, the application and VM-specific predictive
  models can be derived by exploiting various regression and neural
  network based techniques. These models can then be used to predict
  the application's performance using the in-situ profiled resource
  contention with the micro-benchmarks. We evaluated uPredict
  extensively with representative benchmarks from PARSEC, NAS Parallel
  Benchmarks and CloudSuite, on both a private cloud and two public
  clouds. The results show that the average prediction errors are
  between 9.8\% to 17\% for various predictive models on the private
  cloud with high resource contention, while the errors are within 4\%
  on public clouds. A smart load-balancing scheme powered by uPredict
  is presented and can effectively reduce the execution and turnaround
  times of the considered application by 19\% and 10\%, respectively.

\end{abstract}

\maketitle

  
\section{Introduction}\label{sec:introduction}
\input{SecIntroduction}

\section{Closely Related Work}\label{sec:related}
\input{SecRelatedWorks}

\section{VM Performance Profiling}\label{sec:profiling}
\input{SecProfiling}
\section{User-Level Predictive Framework}\label{sec:methodology}
\input{SecMethodology}

\section{Validation and Evaluations}\label{sec:evaluations}
\input{SecEvaluations}

\section{Discussions and Future Work}
\label{sec:limits}

\noindent{\bf Micro-Benchmarks:} \eat{As mentioned in
Section~\ref{sec:evaluations}, }One of the factors affecting the
accuracy of the proposed methodology is that our memory
micro-benchmark only profiles the contention of off-chip memory
resources and has a single memory access pattern. To improve accuracy,
we will investigate micro-benchmarks with smaller working sets to
profile cache contentions as well as other memory access patterns in
the future. Moreover, we will study micro-benchmarks to profile the
contention of network resources so that the proposed methodology can
be extended to network-intensive applications.

\vspace{.1in}
\noindent{\bf Data-Input: }\eat{ Another limitation of this work is not
taking the varying input data for the benchmarks into
consideration. our benchmarks are using the same inputs during
  the training and prediction phases.} The goal of this research is to
investigate the feasibility of predicting cloud application's
performance under resource contention from the ordinary cloud users'
perspective. Hence, we intentionally used the same data inputs for the considered benchmarks to eliminate
the impact from input variations, so that the main cause of
performance fluctuation is resource contention. The evaluation can
thus be focused on the proposed methodology's accuracy to predict the
impacts of resource contention. In our future work, we will extend the
proposed methodology to include predictions under input data
variations. Nonetheless, it is worth noting that many cloud
applications, such as certain machine-learning and data parallel
applications, may be repeatedly executed with similar
workloads~\cite{2012-Agarwal-NSDI,2012-Ferguson-EuroSys,2017-Alipourfard-NSDI}. Therefore,
the methodology presented in this paper should already work well for these
applications.

\vspace{.1in}
\noindent{\bf Long Running Applications: } The evaluation results show
that the predictive models do not work well for long execution times
(outliers) that were not seen in the training data. We plan to
incorporate extreme value theory with Neural Network to improve the
prediction accuracy for these unusual
cases~\cite{2018-Rudd-TPAMI}. Moreover, for long running applications,
they are more likely to experience changes in resource contention
during their executions. We will consider adaptive predictive models
by exploiting periodic profiling techniques in our future work.

\section{Conclusions}\label{sec:conclusions}
\eat{  However, as ordinary cloud users have limited control and
  access to the underlying execution environments, they cannot utilize
  techniques presented by previous studies to predict the performance
  of their applications under resource contention.}

The resource contention in multi-tenant cloud environment can cause
performance fluctuations for cloud applications. Without accurate
knowledge of their performance, it is very difficult for ordinary cloud users to
plan their resource allocations on the clouds. In this paper, we
proposed {\em uPredict}, a user-level profiler-based performance
predictive framework for single-VM applications running in
multi-tenant clouds. First, from the perspective of ordinary cloud
users, {\em uPredict} adopts three specially devised micro-benchmarks
to assess the contention of CPUs, memory and disks, respectively, in a
VM. Then, predictive models based on regression and neural network
(NN) techniques\eat{ to learn the sensitivity of cloud applications to
  resource contention} are developed. The proposed uPredict and the
considered predictive models were evaluated extensively with
representative benchmarks from PARSEC, NAS Parallel Benchmarks and
CloudSuite, on one private cloud server and two public clouds (Amazon
AWS and Google GCE). Our evaluation results show that, even on the
private cloud that has quite high resource contention, the average
prediction errors are between 9.8\% to 17\% for different predictive
models. Here, the NN-based models with hyper-parameters optimization
perform better (about 9\% reduction in prediction errors) than the
regression-based models but with much higher training overheads. For
public clouds that normally have much less contention stable, the
average prediction errors of the considered benchmarks are below
4\%. A use case of {\em uPredict} in load-balancing shows that, the
execution and turnaround times of the considered application can be
effectively reduced by up to 19\% and 10\%, respectively, compared to
the simple queue-based load-balancing scheme.

\eat{
Our results show that under moderate
level of contention in the cloud, regression and SVR model provide
similar accuracy as neural network models. However, NN models will
outperform aforementioned models in terms of accuracy in a more
contentions environment.  We evaluated the proposed methodology with
seventeen benchmarks on a private cloud, Amazon Web Services, and
Google Compute Engine. The evaluation results show that the proposed
methodology is highly accurate with average errors of only 10.3\%,
3.8\% and 3.4\% for each of the three clouds, respectively. These
results demonstrated the feasibility of performance prediction for
ordinary cloud users without the need of knowing or controlling the
underlying execution environment.
}

\newpage
\bibliographystyle{ACM-Reference-Format}
\bibliography{19-sc}

\end{document}

%% file: SecIntroduction.tex
Cloud computing has been adopted by many organizations as their main
computing infrastructure due to its low cost of ownership and flexible
resource management~\cite{2018-RightScale}. However, applications
running on the clouds, especially on the public clouds, usually share
hardware resources with other virtual machines (VMs) and applications
from other cloud users/tenants. Such hardware resource
sharing among multiple tenants causes resource contention, which in
turn degrades the performance of applications running on
clouds~\cite{2016-Leitner-TOIT}. Moreover, the resource contention can
vary due to changes of co-located VMs and their applications, which
makes a target cloud application experience uncontrolled performance
variations and fluctuations at
runtime~\cite{2011-Iosup-TPDS,2010-Ostermann-CC}.


However, to maximize the cost benefits of cloud deployments with
optimal resource allocation~\cite{2011-SC-Mao, 2012-Malawski-SC}, or
to satisfy the timeliness requirements of time-sensitive
applications~\cite{begam-cloud-2018}, cloud users may need to have an
accurate knowledge of the performance of their applications.  For such
a purpose, cloud users need facilities and tools to predict the
performance of their applications under various levels of resource
contention at runtime. While there have been many studies proposed to
predict an application's performance under hardware resource
contention~\cite{christina,
  Mage2018,bubbleUp2011,2013-Tang-ASPLOS,deepdive2013atc,esp2017},
they usually rely on the access to and control over the underlying
execution environment, which makes them not applicable for cloud
users.

Cloud services are typically offered to users as black boxes,
where a user cannot control the cloud execution environment to specify
the set of VMs/applications that should be executed together to share
hardware resources. As a result, it is hard for a cloud user to obtain
an isolated execution environment to profile an application's
contention sensitivity on the cloud service's hardware as did in prior
work~\cite{christina,Mage2018,bubbleUp2011}. Moreover, as cloud users
cannot select the co-runners of their applications, they have to
measure or estimate the severity of resource contention and the
associated impacts on their applications' performance during
execution. Given that cloud users generally have no direct access to
the underlying hardware components and virtual machine hypervisors,
they usually cannot utilize common execution inspection tools used by
prior studies, such as hardware performance monitoring units (PMU), to
obtain accurate estimations on the impact of the
contention~\cite{2013-Tang-ASPLOS,deepdive2013atc,esp2017}.

Therefore, it is imperative to design and develop performance
prediction schemes for ordinary cloud users. Although some recent
studies have addressed this problem, there are still some limitations.
In~\cite{paris}, Yadwadkar et al. developed {\em PARIS}, which
exploits resource profiling information to predict the performance of
an application in a VM when it is deployed on different public cloud
services. Similarly, Scheuner and Leitner employed micro-benchmarks to
test and predict the performance of different types of VMs across
public cloud services~\cite{cloud18}, where a large number of
micro-benchmarks have been deployed. Although these studies can
predict an application's {\em average} performance on various VMs
and/or different cloud services, they cannot be utilized to predict
the {\em in-situ} performance of an application while taking the runtime
resource contention into consideration. An in-situ performance
prediction model enables the users to schedule their
tasks/requests to the VMs that provide the best performance during
execution and thus to improve their quality of services (as
illustrated with the case study in Section~\ref{sub:casestudy}).

In this paper, focusing on single-VM applications, we propose {\bf \em
  uPredict}, a user-level profiler-based predictive framework in
multi-tenant Infrastructure-as-a-Service (IaaS) clouds. Here, to
profile and assess the resource contention of a target VM that is
caused by the colocated unknown VMs and their applications on the same
host, three micro-benchmarks are devised to probe its CPUs, memory and
disks, respectively. Note that, such resource contention can have
various impacts on performance of different applications. To establish
application-specific relationship between its performance and the
profiled resource contention, the micro-benchmarks and an application
are executed sequentially and repeatedly in a given VM to collect
their performance data while the colocated unknown VMs/applications on
the same host may change over time.

With the in-situ profiled resource contention of a VM by the
micro-benchmarks and the measured performance of an application, the
application/VM-specific performance predictive models can be built to
learn the application's sensitivity to the contention of different
resources. In this work, we considered both regression and machine
learning based techniques, including 2-degree polynomial
regression~\cite{zou2005regularization, tibshirani1996regression,
  hoerl1970ridge, sgd2004}, Support Vector Regression
(SVR)~\cite{1998-Gunn-ISISTR} and Neural Networks (NN)
models~\cite{neuralnetwork1992}.\eat{In particular, four regression
  techniques, including Elastic Net, Lasso, Ridges and Stochastic
  Gradient Descent (SGD), were used to train the multivariate 2-degree
  polynomial regression models~\cite{zou2005regularization,
    tibshirani1996regression, hoerl1970ridge, sgd1952,sgd2004}. For
  NN-based models, a fixed NN structure with given number of layers
  and neurons per layer may not work the best for all
  applications. Therefore, two hyperparameter optimization algorithms,
  Random Search and Bayesian Optimization, were employed to
  automatically search for the best hyperparameter for each
  application/VM-specific
  model~\cite{2012-Snoek-NIPS,2012-Bergstra-JMLR}.} The polynomial
models are selected as they are fast to train and may work well if the
relationship between an application's performance and the profiled
data from the micro-benchmarks is indeed polynomial. The SVR and NN
models are selected for cases where the relationships are more complex
than polynomial. Once an application/VM specific predictive model is
derived, the micro-benchmarks can be executed to profile the VM's
resource contention in the current execution environment and the
profiled resource contentiousness can be fed into the model to
predict the application's execution times.

We have evaluated {\em uPredict} extensively using the representative
benchmark applications from PARSEC~\cite{PARSEC}, NAS Parallel
Benchmarks (NPB)~\cite{nas} and CloudSuite~\cite{cloudsuit} in three
different clouds, including a private cloud with OpenStack, Amazon Web
Services (AWS)~\cite{aws-machine} and Google Compute Engine
(GCE)~\cite{google-machine}. First, we validated {\em uPredict} in the
private cloud where the resource contention was introduced in a
controlled manner through changing the number of background VMs and their
applications after each given interval. The predicted performance
using the predictive models for the considered benchmark applications
is shown to be close to and follow the same pattern as the measured
one, which in turn indicates that the micro-benchmarks can effectively
assess the severity of resource contention when the background
VMs/applications change. This illustrates the feasibility of
performance prediction using uPredict for ordinary cloud users without
the need of knowing or controlling the underlying execution
environment.

The prediction errors (i.e., accuracy) of the considered predictive
models were also evaluated. The results show that, for the considered
applications and VMs, the NN-based models are generally more accurate
with the average prediction errors being 9.8\%, 3.8\% and 3.4\% for
the private cloud, AWS and GCE, respectively. In comparison, the
polynomial regression and SVR models perform slightly worse in the
private cloud that has high resource contention with the
average prediction errors being 17\% and 13\%,
respectively. However, on AWS and GCE where the level of contention is
lower than our private cloud, the polynomial regression models and SVR
models have almost the same prediction errors on average compared to
those of NN-based models. However, we would like to point out that,
the NN-based models require hyperparameter optimizations, which can
introduce larger training overheads.

As an application of {\em uPredict}, a case study on load-balancing
for two VM servers on two different host machines was presented. For
comparison, we considered a simple queue-based load-balancing scheme
that makes load distribution based on the queue length (i.e., the
number of requests) on each VM server without considering their
resource contention. The results show that the {\em uPredict} based
load-balancing can achieve about 19\% reduction in average
application execution times and 10\% reduction in their turnaround
times.

\eat{
However, we would like to point out that NN-based models require
considerable tuning of its hyper-parameters (e.g., the number of
layers and neurons per layer) to achieve the reported level of
accuracy. Poorly chosen hyper-parameters can considerably worsen the
accuracy of NN-based models (as high as 155\% prediction error).\eat{
  This hyper-parameter issue does not only affect accuracy, it may
  also prevent the reproduction of our predictive models.} Therefore,
to provide best prediction results with NN-based models, we employed
two different hyper-parameters optimization algorithms,
HyperOpt~\cite{hyperopt} and
Scikit-Optimization~\cite{scikitoptimize}, to automatically search for
the best hyper-parameters for each application-specific NN-based
model. Moreover, the hyper-parameters optimizations can significantly
increase the training time (from 20 minutes to 2 hours), given the
large number of configurations for the NN structures to be
evaluated. Therefore, considering their low training overheads and
relatively high prediction accuracy, the polynomial regression models
and SVR models may be more practical than NN-based models on public
clouds (e.g., AWS and GCE).
}

The main contributions of this work are summarized as follows: 
\begin{enumerate}

  \item A user-level profiler-based predictive framework, {\em
    uPredict}, is proposed, which aims at providing accurate
    performance prediction of single-VM applications for ordinary
    cloud users in multi-tenant cloud environment without the
    knowledge and controlling of co-located VMs and their
    applications;

  \item Three micro-benchmarks are specially devised to profile and
    assess the contention of CPUs, memory and disks of a VM,
    respectively; Using the in-situ profiled resource contention data,
    both regression and neural network (NN) based techniques are
    exploited to build application/VM-specific performance predictive
    models;

  \item The proposed framework, micro-benchmarks and various
    predictive models are evaluated extensively with the
    representative benchmark applications from several benchmark
    suites in both private and public clouds; A case study of
    utilizing {\em uPredict} in a smart load-balancing scheme was also
    investigated; The evaluation results show the feasibility and
    effectiveness of {\em uPredict} for ordinary cloud users.

\eat{
\item A highly-accurate resource-contention-aware performance
  prediction methodology from an ordinary cloud user's perspective
  without the need of knowing or controlling the underlying execution
  environment.

\item An extensive evaluation of different types of machine learning
  models and their corresponding effects on prediction accuracy under
  different levels of contention. In particular, hyper-parameters
  optimization was incorporated into our methodology to find the best
  NN models and achieve high research reproducibility.

\item A thorough evaluation of the performance prediction methodology
  with seventeen benchmarks on three clouds, demonstrating the
  feasibility of such user-oriented predictions on public clouds.
}

\end{enumerate}

The rest of the paper is organized as follows.
Section~\ref{sec:related} reviews closely related
work. Section~\ref{sec:profiling} discusses the micro-benchmarks that
are devised to profile and assess the resource contention of a VM in
multi-tenant environment. Section~\ref{sec:methodology} presents the
proposed {\em uPredict} framework and several predictive models. The
experimental setups and evaluation results are discussed in
Section~\ref{sec:evaluations}. Section~\ref{sec:limits} points out the
limitations of this study and our future
works. Section~\ref{sec:conclusions} concludes the paper.



%% file: SecRelatedWorks.tex
\textbf{Cloud Performance Variation Analysis}. Many research studies
have observed and analyzed the performance variations of cloud
applications~\cite{12d,19d,7est}. Iosup et al. were among the first
reporting the performance variations in public
clouds~\cite{2011-Iosup-TPDS}. The same research group also conducted
an in-depth analysis of the performance variation on production public clouds~\cite{2010-Ostermann-CC,2011-Iosup-CCGrid}. Leitner and Cito
conducted a performance analysis on multiple public clouds and
observed that the resource contention is a major cause for single VM
performance variations~\cite{2016-Leitner-TOIT}.
In a more recent study, Maricq et al. also presented large amount of
data, suggesting the extensiveness of performance fluctuation in the
clouds~\cite{2018-Maricq-OSDE}.
Our work is inspired
by these studies on cloud performance variations.

\vspace{.05in} \textbf{Contention-aware Performance Prediction from
  Cloud Service Provider's Perspective}. There has been a flurry of
research works on predicting application performance under resource
contention from the perspective of cloud service providers and data
center operators.  Paragon is a heterogeneity and interference-aware
data center scheduler~\cite{christina}. To make scheduling decisions,
Paragon profiled its applications in a controlled environment to
determine their contentiousness and sensitivity to contention.  Quasar
estimated the resources that a data center application required to
meet its QoS goals by profiling its performance on specific hardware
running with specific micro-benchmarks~\cite{2014-Delimitrou-ASPLOS}.
The idea of Paragon and Quasar was later extended to consider multiple
levels of hardware heterogeneity in data centers~\cite{Mage2018}.
Bubble-up characterized the sensitivity of a data center application
and predicted the application's performance under contention by
injecting pressure into the memory system.~\cite{bubbleUp2011}.
Bubble-flux dynamically injected pressure into the memory system to
measure the application's instantaneous sensitivity to contentions
using readings from hardware Performance Monitoring Units
(PMUs)~\cite{2013-Yang-ISCA}.  Govindan et al. designed synthetic
benchmarks to clone the cache behaviors of a set of applications,
which were later used to profile and predict the performance
degradation when two applications running together~\cite{cuanta2011}.
Q-Clouds dynamically adapted the resource allocations for co-running
VMs based on the run-time performance obtained from
PMUs~\cite{qcloud2010eurosys}. ESP predicted the performance impact of
contentions for a known set of applications using
regularization~\cite{esp2017}. Oktopus improved the performance
predictability of a cloud system by offering virtualized network
interfaces~\cite{2011-Ballani-SIGCOMM}. 

These cloud-provider-side prediction methodologies typically required
the control of the execution environment in their profiling phases,
including directly specifying the hardware platforms and the
co-running tasks used in the profiling. Many of these studies also
assumed a known set of applications that might be executed and/or
required accesses to low-level hardware PMUs. Our work, however, aims
at performance prediction for ordinary cloud users who have no control
of the execution environments, no knowledge of the co-running
applications and no access to the hardware PMUs.

\vspace{.05in}
\textbf{Performance Prediction from User's Perspective}. There were
also studies on cloud performance prediction techniques for cloud
users. Scheuner and Leitner employed micro-benchmarks to test and
predict the performance of different types of VM instances across
public cloud services~\cite{cloud18}. They considered 23 different
micro-benchmarks and validated the methodology against two
applications. The authors also observed that not all micro-benchmarks
were necessary for performance prediction. uPredict, however, employed
a much smaller but clearly defined set of micro-benchmarks and was
validated against 17 benchmarks with various behaviors.

PARIS predicts the performance of an application when it is deployed
on different types of cloud instances~\cite{paris}. PARIS did not
consider the impact of resource contention and experienced up to 50\%
RSME (Root Mean Squared Error). Li et al. predicted the performance of
cloud applications when they were allocated with a different number of
CPUs~\cite{NICBLE2017}. Ernest built performance models based on the
behavior of the job on small inputs and then predicted the performance
on larger data sets~\cite{ernest2016nsdi}. Baughman et al. employed
online and offline profiling to predict an application's performance
when deployed on certain VM types with certain
inputs~\cite{2018-Baughman-UCC}. Mariani et al. proposed to let cloud
service providers build performance models to help user predict the
performance of High-performance Computing (HPC) applications running
on different VM types~\cite{2017-Mariani-CCGRID}.  Wolf et
al. proposed a method to automatically model the performance of HPC
applications from limited profiling data to identify scalability
bottlenecks~\cite{2016-Wolf-SPPEXA}. PRIONN is an automated run time
and I/O usage prediction tool for HPC clusters~\cite{2018-Wyatt-ICPP}.
Zhai et al. investigated large-scale HPC application performance
prediction with deterministic replay~\cite{2016-Zhai-IEEEComputer}.
Clemente-Castello et al. proposed a methodology to predict MapReduce
application's performance in hybrid clouds~\cite{2018-CLemente-TPDS}.
Friese et al. presented a novel hierarchical critical path analysis
methodology to predict the performance of irregular
applications~\cite{2017-Friese-IPDPS}.  Jiang et al. proposed a model
to predict whether a type of VM should be provisioned for a specific
tier (e.g., web server and database) of a web
application~\cite{2011-Jiang-Webapps}. This model, however, does not
consider the impact of resource contention. Farley et al. proposed a strategy
to estimate the future performance of a VM based on the observed
performance or resource usages~\cite{2012-Farley-SoCC}.\eat{ However, the
strategy did not aim to predict the exact performance.}

Unlike uPredict, these studies did not intend to predict the performance of
cloud applications under currently observed level of resource contention in
multi-tenant clouds.

%% file: SecProfiling.tex
\begin{figure*}[htbp]
  {\includegraphics[width=1\textwidth]{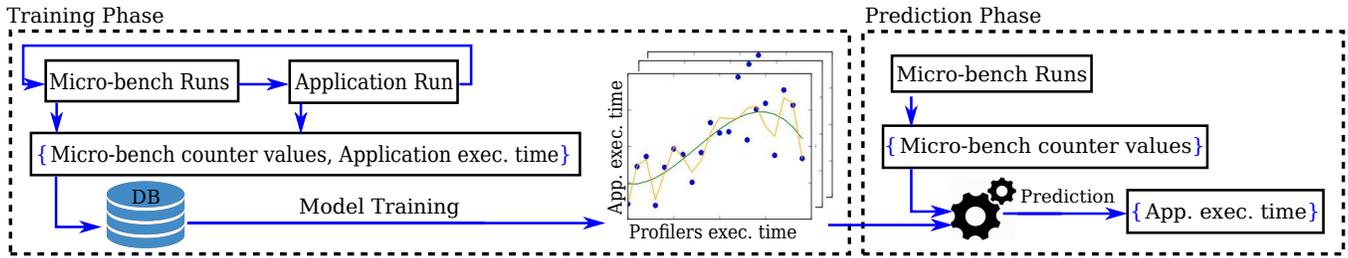}}\\
  \vspace{-.1in}
  \caption{Overview of uPredict and its workflow for performance modeling and prediction.}
\label{fig:overview}
\end{figure*}

In a multi-tenant cloud environment, a VM normally shares and contends
for the underlying hardware resources with other colocated VMs and
applications. The deliverable performance to user applications by a VM
depends heavily on resource contention in CPUs, memory and disks,
especially for single-VM applications with limited network
activities. Given that ordinary cloud users do not have access to
hardware performance counters in the hypervisor and underlying
hardware of the host machine, we focus on user-level profiling
techniques to obtain the in-situ resource contention of a VM. Note
that, although cloud users can retrieve the utilization of various
(virtual) resources from a VM, such information does not reflect
resource contention due to other colocated VMs and thus cannot
indicate its actual performance. For example, a reading of 100\% CPU
utilization from a VM just indicates that all its virtual CPUs are
fully utilized by user applications without any insight regarding to
the actual deliverable performance for the applications.

Therefore, to infer the perceivable performance from the perspective
of cloud user applications regarding to the resource contention of a
VM, we employ several micro-benchmarks to assess its resource
contention due to interference from other collocated VMs and
applications on the same host machine.  Intuitively, the slowed (or fastened) progress of a micro-benchmark reflects the increased
(or decreased) contention for the corresponding resource. 
There are many micro-benchmarking tools available. Some are designed by vendors (e.g., Intel's VTune Amplifier~\cite{intel-vtune}). Some are open-source benchmarks (e.g., lmBench3~\cite{lmbench} and iozone). However, we decide to design and implement our own set of micro-benchmarks because of the limitations with existing micro-benchmarks and tools. Tools from the vendors may use Performance Monitoring Unit (PMU) counters (e.g., Intel VTune) to analyze the application behavior (e.g. cache miss). These tools are not applicable for cloud users, as cloud users do not have accesses to these low-level performance counters. Open-source benchmarks have restricted behaviors, such as the execution length and memory access patterns. Therefore, open-source benchmarks cannot be used in their original form.  Because of these limitations, we implemented our own micro-benchmarks for the profiling. Our micro-benchmarks are based on the open-source benchmarks. However, they are designed to allow variable execution times (and thus variable profiling execution lengths) and variable memory/storage memory access patterns. By experimenting with different execution times and access patterns, it is possible to find the best profiling methodology empirically. 

Prior work
has shown that resource contention mainly happens in CPUs, memory,
storage and network resources~\cite{christina}. Note that, for
single-VM applications, the impact of network contention is
negligible. Hence, we designed three micro-benchmarks for uPredict to
probe the contention of CPUs, memory and disks, respectively.  In the following paragraphs, we explain the detailed design of our micro-benchmarks.


\vspace{0.05in}
\subsection{Micro-benchmarks}\label{sub:microbench}

\noindent{\bf CPUs:} For contention in CPUs, a multi-threaded
micro-benchmark is designed to stress the performance of the virtual
CPUs of a given VM. Here, each thread maintains an {\em in-register
  counter} that is initiated with zero. During execution, the
thread repeatedly increments the counter for a fixed amount of profiling execution time specified by the user. Such in-register operations ensure that the micro-benchmark's
performance is not affected by memory at runtime and thus examine the
contention in CPUs to the maximum extent. The number of threads
deployed in this micro-benchmark will be equal to the number of
virtual CPUs of the target VM. The total number of increment operations carried out by all threads for the in-register counter will be recorded. In the end, this number ($c_{cpu}$) will be used as the indicator of the progress of this benchmark and the contention level for the virtual CPUs in the target VM.

\vspace{.05in}
\noindent{\bf Memory:} Similarly, the memory micro-benchmark will try
to stress the memory bandwidth of the target VM to the maximum
extent. This micro-benchmark accesses a 2GB array with a
stride of 128 byes. The objective of such a memory access
pattern is to ensure that each data access is issued to off-core
memory rather than the caches. Again, the number of threads in
the micro-benchmark equals to the number of virtual CPUs in the VM and
each thread will access an equal portion of the array and increases the local counter by one for each access. The total number of memory accesses by all threads in a specific amount of profiling time for this micro-benchmark ($c_{mem}$) will provide us the insight into the
memory contention and its impacts on performance experienced by user
applications in the target VM.

\vspace{.05in}
\noindent{\bf Disk I/Os:} For I/O performance of the target VM, we
design the disk micro-benchmark that reads 256MB data from the VM's
disk with the page size of 4KB, and by each disk access the local counter value will be incremented. During the execution of this
micro-benchmark, the OS file cache should be disabled to all file operations access the disk. Four threads are use for this
micro-benchmark to fully exercises the disk without causing too much internal I/O contention. Again, the total number of disk access operations within a specific profiling time for this micro-benchmark ($c_{disk}$) is use to
assess the contention level of the VM's disk operations. 

These micro-benchmarks will be invoked sequentially right before
the execution of a user's application to get the in-situ resource
contention for the respective resources.
\subsection{The Length of Profiling Executions}\label{sec:duration}

Although it is desirable to reduce profiling overhead, short executions of the
micro-benchmarks may not be able to completely capture the actual severity of resource
contention. For majority of the experiments conducted in this paper, we executed each micro-benchmarks for 3 seconds to ensure the actual severity of contention was properly captured. In Section~\ref{sec:profiling_sensitivity}, we conducted a sensitivity test on how the length of profiling affects uPredict's accuracy. The sensitivity test shows that 3 seconds indeed can provide accurate profiling results, while lower profiling length may also suffice.

%% file: SecMethodology.tex
In this section, we first present an overview of {\em uPredict: a
  user-level profiler-based predictive framework}. Then, based on
regression and neural-network techniques, several predictive models
are discussed that have different complexities and capabilities.

\subsection{Overview of {\em uPredict}}

Figure~\ref{fig:overview} shows the overview of {\em uPredict} and
illustrates the workflow of performance modeling and prediction for an
application running on a VM in a multi-tenant cloud. There are two major phases in {\em uPredict}: the {\em
  training} and {\em prediction} phases. The first step in the
training phase is to collect the training performance data. Here, in
each iteration, the three micro-benchmarks are executed first for a fixed amount of time and their access counter values being denoted as $\{c_{cpu}, c_{mem},
c_{disk}\}$ to assess the in-situ contentiousness of CPUs, memory and
disks of a VM, respectively. Then, the target application is
executed right after the micro-benchmarks with its execution time
being denoted as $t_{app}$. The access counter values of
micro-benchmarks and execution times of the application will form a data tuple $\{c_{cpu},
c_{mem}, c_{disk}, t_{app}\}$, which represents the implicit
relationship between the application's performance and the
profiled resource contention by the micro-benchmarks.

A set of training data tuples needs to be collected by repeating the above
process for the target application and VM, where the resource
contention from other VMs and applications on the same host machine
can vary. The data tuples can then be used to train various
application and VM specific performance predictive models based on
different regression and neural network techniques, as discussed
next. The number of data tuples in the training set can affect the
accuracy of the derived models and the trade-offs are evaluated in
Section~\ref{sec:evaluations}. The second phase of performance
prediction utilizing the derived predictive models is detailed in
Section~\ref{sec:prediction}.

\subsection{Predictive Models in {\em uPredict}}
\label{sec:predictive-models}

The key step in the training phase is to learn the relationship
between the application execution time and the resource contention
represented by the micro-benchmarks' access counter values from the collected
data tuples and derive a predictive model. That is, the parameters of
the function $f$ in Equation~(\ref{eq:reg_model}) need to be learned.\
\begin{equation}
  \label{eq:reg_model}
  t_{app} = f(c_{cpu}, c_{mem}, c_{disk})
\end{equation}
where the exact form of the function $f$ and its parameters depend on the specific
regression or machine-learning technique being used.

In other words, we use the micro-benchmark access counter values as the
features to predict the execution time of the target
application. Intuitively, the access counter values of micro-benchmarks are
the indicators of contention severity of different resources, which
are in turn used to predict the execution time of the application as
shown in Section~\ref{sec:prediction}. Note that, the execution time
of an application may also depend on its input data. In this work, we
assume that an application's execution time is affected only by the
resource contention from other collocated VMs and their applications
on the same host machine, where the input data for the application in
each execution has the same or similar size. It has been shown that
many recurring cloud applications are indeed repeatedly executed with
similar
workloads~\cite{2017-Alipourfard-NSDI,2012-Agarwal-NSDI,2012-Ferguson-EuroSys}. Moreover,
the developed predictive models can be easily extended to consider an
application's input data, especially when such data size has a known
(e.g., linear) relation with the application's execution time.

Moreover, as applications have different behaviors and sensitivities
to resource contention when running in a given VM, a single predictive
model may not perform well for all applications. Therefore,
application and VM specific predictive models will be derived for each
application running on the considered VM\eat{in {\em uPredict}
  based on the various modeling techniques}. In what follows, the
details of several regression and neural network based modeling
techniques and their training processes are presented, which generally
have different complexities (thus overheads) and prediction accuracies
as shown in Section~\ref{sec:evaluations}.

\subsubsection{Polynomial-Regression based Models}

We first considered polynomial-regression based predictive models,
which usually take a short amount of time to train and
predict. Consequently, if polynomial models can provide good prediction accuracy for an application running in a VM,
there is no need to employ other more expensive and heavy-headed
machine-learning models. To ensure that polynomial-regression models
are thoroughly evaluated, we explored four regression (model training)
techniques, which are Elastic Net
Regularization~\cite{zou2005regularization}, Lasso
Regression~\cite{tibshirani1996regression}, Ridge
Regression~\cite{hoerl1970ridge}, and Stochastic Gradient Descent
(SGD)~\cite{sgd2004}. The exact parameters for training these models can be found in Section~\ref{sec:evaluations}. Additionally, experiments have been conducted
with linear regression and 3-degree polynomial regression. However,
for the considered benchmark applications, they perform inferior
comparing to 2-degree polynomial models. Therefore, we only report the
results of 2-degree polynomial predictive models in this paper.


\subsubsection{Support Vector Regression (SVR) based Models}


We also considered Support Vector
Regression (SVR) based models~\cite{1998-Gunn-ISISTR}, which may
potentially provide higher accuracy than polynomial models but with larger
training and prediction cost. 
SVR is based on the popular machine-learning classifier, Support
Vector Machine (SVM), with the introduction of an alternative loss
function (in our case, the popular epsilon-insensitive
function)~\cite{1998-Smola-Algorithmica}. The main benefit of SVR is
that it allows us to build more complex and non-linear models within
reasonable amount time, as an application's behaviors running in the
clouds may not always be expressible with polynomial or linear
equations of resource contention~\cite{1998-Gunn-ISISTR}.

For SVR, the function $f$ has the format as shown in
Equation~(\ref{eq:svr_model}), where $x$ represents the profiled
execution times from the micro-benchmarks (i.e., $x$ is the tuple
$\{c_{cpu}, c_{mem}, c_{disk}\}$).  The $l_i$ in
Equation~(\ref{eq:svr_model}) represents the micro-benchmark execution
time tuple from a training sample.  The training process will
determine the actual values of all $\omega$'s and the $b$ with a
predefined kernel function $K$.

\begin{equation}
  \label{eq:svr_model}
  \begin{split}
    t_{app}=&\sum\omega_i\cdot K(x, l_i) + b
  \end{split}
\end{equation}

We employed the SVR implementation from Scikit-Learn~\cite{scikit}
with the default Gaussian kernel, which can be expressed as,
\begin{equation}
  \label{eq:svr_kernel}
  \begin{split}
    K(x, l_i) = exp(-\gamma ||x-l_i||^2)
  \end{split}
\end{equation}

The value of $\gamma$ is also automatically determined by the
Scikit-Learn implementation by default.

\subsubsection{Neural Network (NN) based Predictive Models}\label{sec:nn_models}
In addtion to SVR, we also considered Neural Network (NN) based
models~\cite{neuralnetwork1992,1991-Specht-TNN}.  In uPredict, NN
models are configured to conduct regression analysis. These models are
more generic than SVM and can approximate nearly any function,
potentially allowing {\em uPredict} to model any behaviors of an
application running in a VM with higher training
costs~\cite{1989-HORNIK-NN,1995-Bishop-NeuralBook}. However, we have
observed that the accuracy of NN models can be significantly affected
by their structures, that is, the number of layers and the number of
neurons in each layer. Even when the same set of training data is used,
the worst NN structure can have the prediction error to be more than 10 times
higher than the best one. Therefore, training NN models with good
accuracy is not simply just feeding the training data into a model, it
also involves optimizing the structures of the NN models. Moreover,
the best NN model structure also varies for different applications and
VMs, implying that {\em uPredict} methodology needs to individually
optimize the model structure for each pair of application and VM. This
optimization process should be automated so that ordinary cloud
users, who do not have expertise in machine learning, can apply
uPredict to a new application and/or VM.

To automatically optimize NN structures, we employed hyperparameter
optimization techniques, including Tree-structured Parzen Estimator
(TPE) approach and Bayesian
Optimization~\cite{2012-Snoek-NIPS,2011-Bergstra-NIPS}. Hyperparameters
refer to the NN parameters defining the number of layers and the
number of neurons in each layer of a NN model. Both optimization
techniques conduct a search in the optimization search space of the NN
models to find a structure with good accuracy. This search space
defines the maximum number of layers and neurons per layer that can
be used when training NN models. The TPE technique explores the search
space using a tree structure following the accuracy distribution
obtained from previously sampled points in the search space. The
Bayesian Optimization searches for the high-accuracy NN structures
through the Gaussian Process (GP), which is a non-linear
regression technique. Bayesian optimization uses GP to build a
regression model with the already explored NN structures and their
accuracies. The regression model is then used to predict a potentially
better NN structure until a fixed number of NN structures are searched.

Clearly, the accuracy of the NN models in {\em uPredict} depends on the
definition of the search space. It is commonly recommended that the
number of neurons per layer is typically ``no more than 1/30 of the
number of training cases''~\cite{NNFAQ}. As our training sets only contain
up to 1,000 data samples for each benchmark application in the given VMs,\eat{
each layer may need no more than roughly 34 neurons following this
recommendation.} we set the maximum neurons per
layer to be 35 in {\em uPredict}. As neural networks with two hidden layers (four total
layers) can be fully general, we define the maximum number of layers
of uPredict's NN models to be 5 ~\cite{1992-Sontag-TNN}. The extra
layer is added to accommodate the cases where the maximum number of
neurons defined above is not large enough. In summary, uPredict
employs a NN structure search space of maximum 5 fully-connected
layers and maximum 35 neurons per layer. The final optimized NN models
for different applications may not have the same number of layers and
neurons at each layer.

Note that,\eat{ given that we only have up to 1,000 data samples per
  training set,} in uPredict, the hyperparameter optimization for NN
models is applied on the training data sets themselves instead of
separate cross-validation data sets. Theoretically, training and
optimizing on the same data set may lead to over-fitting and thus low
prediction accuracy. However, our experiment results show that using
the same data set for optimizing the NN models can still provide high
accuracy for predicting the performance of the considered benchmark
applications and VMs on the multi-tenant clouds.

\subsection{Performance Prediction in {\em uPredict}}
\label{sec:prediction}


In the prediction phase, before running an application, the
micro-benchmarks are first executed sequentially to profile the
contentiousness of a VM's CPUs, memory and disks in the current
execution environment, respectively. Their access counter values, $c_{cpu}$,
$c_{mem}$ and $c_{disk}$, are then fed into a trained model $f$ for
the application to predict its execution time. Here, the profiled
resource contention by the micro-benchmarks is used to estimate the
contention to be experienced by the application.

For the benchmark applications used in our evaluations, they typically
take less than one hour to execute and our observation shows that the
resource contention is less likely to change significantly within such
a short period of time, especially on the public clouds. However, when
an application does experience a change in resource contention during
its execution, the prediction accuracy of the derived models can be
negatively affected with much higher prediction errors as shown in the
evaluation results (see Section~\ref{sec:evaluations}).

For long running applications, they will be more likely to experience
changes in resource contention during their executions and a
periodically re-profiling technique may be deployed to catch such
changes. However, exploring such a periodic re-profiling option would
require significant modifications to the model building and prediction
process (to consider, for instance, re-profiling intervals), which
is well beyond the scope of this paper and will be investigated in our
future work.


%% file: SecEvaluations.tex
The proposed {\em uPredict} with the aforementioned predictive models
have been evaluated extensively using the representative benchmarks
from several benchmark suites on different clouds. Here, we first
present the experiment setups and explain the data collection
process. Then, the validation of {\em uPredict} on a private cloud
with controlled reource contention is discussed. By considering both a
private cloud and two public clouds that represent different severity of
resource contention, the prediction accuracies (i.e., errors) of the
predictive models are evaluated. Lastly, we present a use case of {\em
  uPredict} in a smart load-balancing scheme on two cloud servers.

\subsection{Experiment Setups}
\label{sub:setup}

\textbf{Representative Benchmark Applications:} A total of 17
benchmarks from PARSEC~\cite{PARSEC}, NAS Parallel Benchmarks
(NPB)~\cite{nas} and CloudSuite~\cite{cloudsuit} have been considered
in our evaluations. Eight of them are from PARSEC, including {\em
  streamcluster}, {\em blackscholes}, {\em bodytrack}, {\em canneal},
{\em facesim}, {\em ferret}, {\em swaptions} and {\em dedup}. For
these PARSEC benchmarks, their native inputs were used in the
experiments. Five are chosen from NPB, which are {\em ua}, {\em lu}, {\em
  sp}, {\em ep} and {\em bt}, and they used the class C data inputs. The
other four are from CloudSuite, including {\em In-Memory Analytic},
{\em Graph Analytic}, {\em Web Search} and {\em Data Serving}. Here,
the large data inputs were used for {\em In-Memory Analytic}, while
{\em Graph analytics}, {\em Web Search} and {\em Data Serving}
benchmarks used the default data inputs. 

These 17 benchmarks are representative and cover a wide range of
applications running in various clouds.\eat{ For instance, the benchmarks
from PARSEC are representative for batch processing jobs, NPB
benchmarks are representative for HPC applications and the ones from
CloudSuite are representative for business applications.} In each
benchmark suite, the selection of these benchmarks is a combination of
technical difficulties (e.g., compilation problems), benchmarks'
resource requirements (needs of multiple VM instances or more memory) and
budget limitation to run the costly experiments on AWS and GCE
clouds. For all the 17 selected benchmarks, sixteen worker threads
were created in their executions.

\vspace{.05in}
\noindent\textbf{Clouds and VM Configurations: } First, for the
private cloud, we utilize an Ubuntu 16.04 server with two Intel Xeon
E5-2630 processors (for a total of 16 cores) and 128GB memory that has
OpenStack Ocata installed. Given that the selected benchmarks include
parallel and data/graph analytic applications, we created a VM of 16
VCPUs and 16GB memory on OpenStack to execute these
benchmarks. Moreover, to introduce resource contention into the
private cloud, up to seven (7) background VMs (with the same VCPU and
memory configuration as the target VM) were randomly created at
runtime, which executed either CPU- or memory-intensive synthetic
applications from iBench~\cite{ibench}. The background VMs and their
applications change after each fixed interval (e.g., 2 hours).

For public clouds, we considered both AWS EC2 and Google Cloud Engine
(GCE). In AWS EC2, we used a single VM of type {\em m5d.4xlarge} to
execute the selected benchmarks. Here, {\em m5d.4xlarge} is the latest
general purpose VM instance with 16 CPUs and 64GB
memory~\cite{aws-machine}. The VM is configured to use an 80GB
standard EBS SSD drive. We used non-dedicated VM instances so that
background VMs and their applications were managed by AWS and were
unknown to us. For GCE, we used a single VM of type {\em
  n1-standard-16} to execute the selected benchmarks, where {\em
  n1-standard-16} is the standard VM instance with 16 VCPUs and 60GB
of memory~\cite{google-machine}. The VM is also configured with an 80GB
SSD drive. Again, the background VMs and their applications were
managed by GCE and were unknown to us. Both the selected VM types in
the public clouds closely match CPUs and memory of the VM in our
private cloud. For all the experiments on the three clouds, we used
Ubuntu Server 16.04 as the OS for the created VMs.

Here, the private cloud was empolyed with two objectives: First, to
validate the correctness of {\em uPredict} with controlled resource
contention from the background VMs and applications; Second, to
evaluate the prediction accuracy of the studied predictive models in
{\em uPredict} under scenarios with extremely high resource
contention, which were usually not seen in the experiments on public
clouds. On the other hand, the experiments on the VMs in AWS EC2 and
GCE were designed to evaluate the effectiveness of {\em uPredict} and
its predictive models. We would like to evaluate whether they can indeed
predict the performance of cloud applications with high accuracy from
the perspective of ordinary cloud users on commercial public clouds
with unknown colocated VMs and workloads.

\eat{
The predictive models were evaluated with eight
PARSEC, five NAS Parallel and four CloudSuit Benchmarks on three
clouds, including a private cloud in our lab, Amazon Web Services
(AWS) and Google Compute Engine (GCE). The experiments on public AWS
EC2 and GCE were designed to evaluate if the proposed predictive
models can indeed predict performance with high accuracy from ordinary
cloud user's perspective on public clouds infrastructure too.

}

\begin{figure*}
\begin{center}
 $\begin{array}{cc}
 \epsfxsize = 3.5in
  {\includegraphics[width=0.49\textwidth]{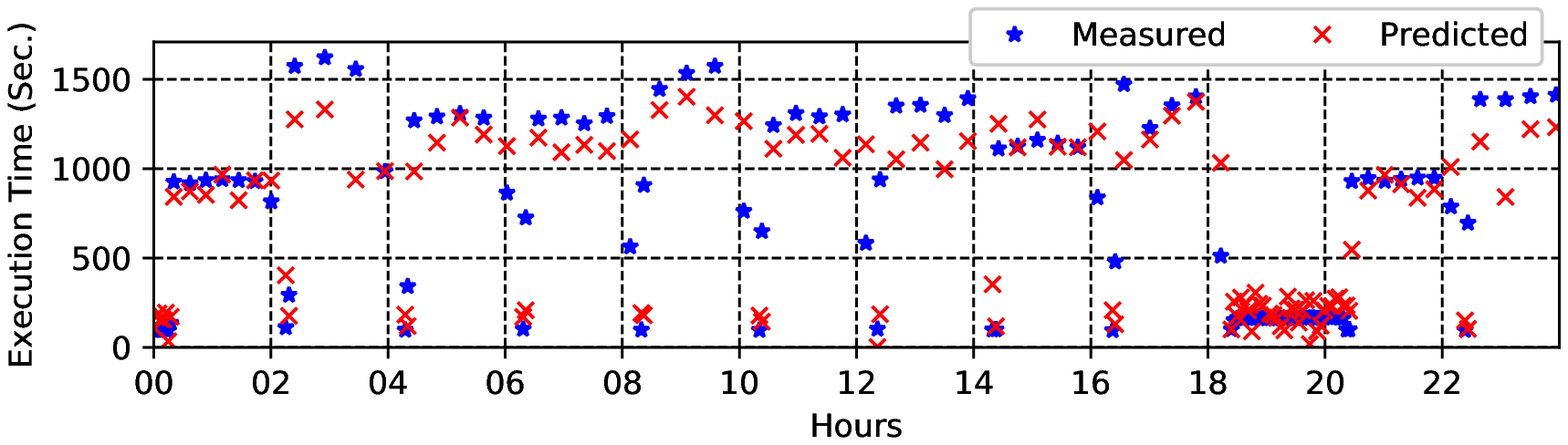}}
&\epsfxsize = 3.5in
  {\includegraphics[width=0.49\textwidth]{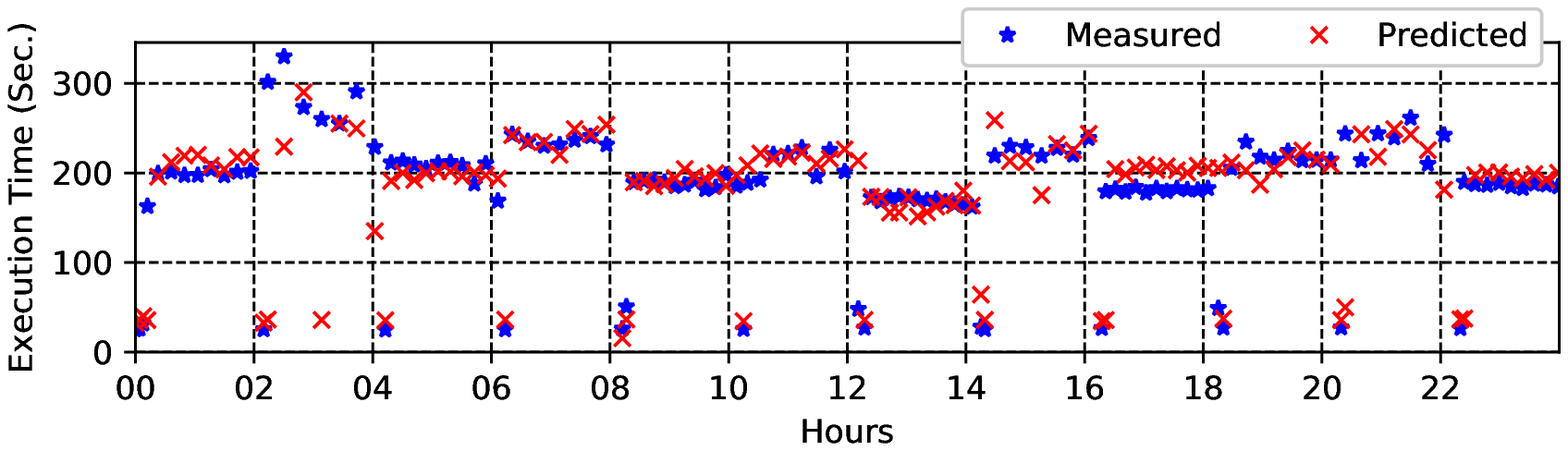}}\\
 \epsfxsize = 3.5in
  {\includegraphics[width=0.49\textwidth]{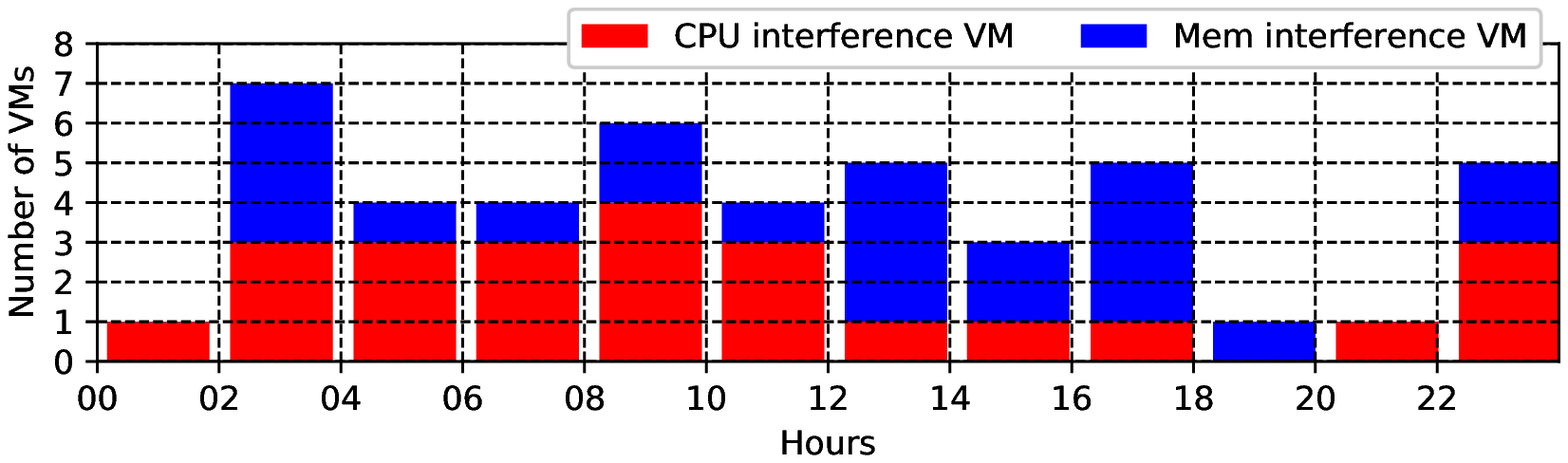}}
&\epsfxsize = 3.5in
  {\includegraphics[width=0.49\textwidth]{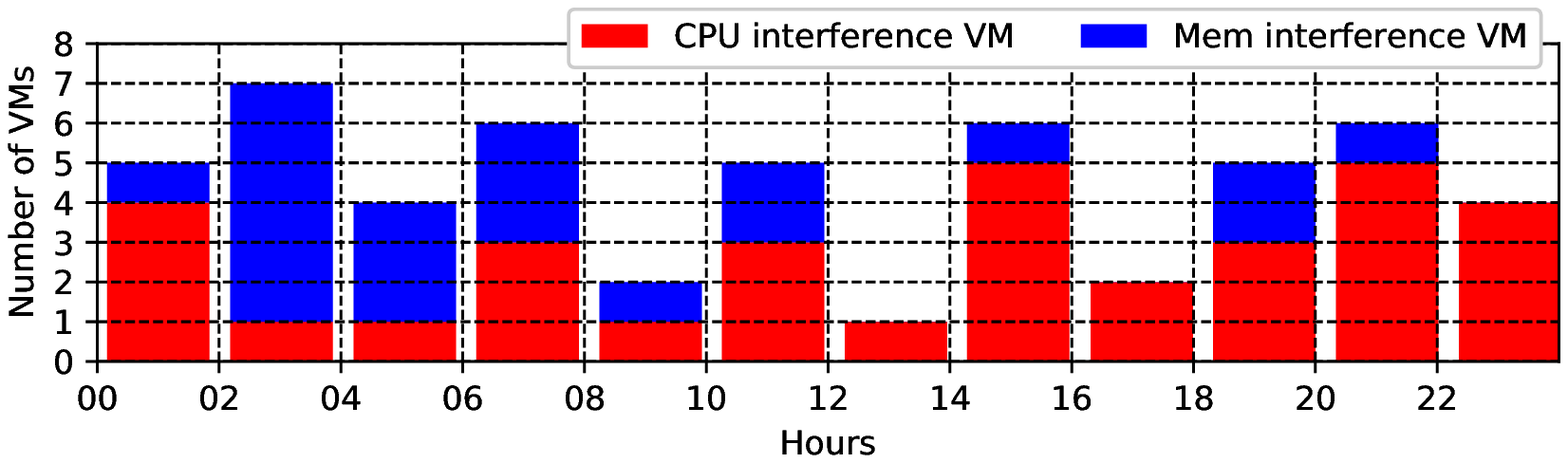}}\\
  \mbox{{\small a. Streamcluster }} 
& \mbox{{\small b. Canneal }} \\
\end{array}$
\end{center}
\vspace{-.1in}
\caption{Measured and predicted execution times for {\em Streamcluster} and
  {\em Canneal} with the background VMs on the private cloud.}
\label{res:validation}
\end{figure*}

\vspace{.05in}
\noindent\textbf{Data Collection: } As illustrated in
Figure~\ref{fig:overview}, a key step in the training phase of {\em
  uPredict} is to collect experimental data regarding to the execution
times of the benchmark applications and those for the
micro-benchmarks.\eat{ Although it is desirable to reduce profiling
  overhead, the short executions of the micro-benchmarks can be
  affected by the VM's scheduling activities and may not be able to
  properly catch the correct level of resource contention. We have
  evaluated the effectiveness of these micro-benchmarks for executions
  from 1 to 10 seconds. The results show that, to get reasonable
  prediction accuracy for our predictive models, each micro-benchmark
  needs to run for at least 3 seconds to obtain the necessarily
  accurate contention measurements.} From Section~\ref{sec:profiling},
the three micro-benchmarks are invoked sequentially right before the
execution of a user's application to get the in-situ resource
contention, which has the total profiling overhead of roughly 9
seconds in each iteration. For the private cloud, with up to seven (7)
background VMs and associated applications, we have run the 17
benchmark applications individually with the micro-benchmarks for a
total of roughly 70 days. For each benchmark application, more than
1,000 data points have been collected and the first 1,000 were used in
the evaluations.

For the PARSEC benchmarks that have relatively short execution times,
we have executed them for about 10 days on both AWS and GCE to collect
more than 1,000 data points in each cloud setting. Again, the first
1,000 were used in the evaluations. For the benchmarks in NPB and
CloudSuite that take more time for executions, we run them for about
20 days on both AWS and GCE, where 777 and 688 data points have been
collected for each of these benchmark applications on the two clouds,
respectively, and all these data points were used in the evaluations.

For each benchmark application, 80\% of the collected data points are
utilized as training data to derive its regression and neural network
based predictive models. The remaining 20\% data points are used as
testing data to evaluate the prediction accuracies (i.e., errors) of
the derived predictive models. Here, instead of designating a fixed
portion of 80\% data points as the training data that may not reflect
the same severity of resource contention experienced by the benchmark
applications for the other 20\%, we adopted a sampling technique to
select the training data points. Specifically, for every 5 consecutive data points
that are more likely to encounter similar resource contention, the
first 4 of them are chosen as the training data, while the last data
point is used as the testing data.

\noindent\textbf{Implementation and Training of the Predictive
  Models:} We used the scikit-learn version 0.19.2
library~\cite{scikit} to implement the four different 2-degree
polynomial regression models and the SVR model. In particular, for the
{\em ElasticNet} and {\em Lasso} algorithms, we used an alpha of 1 as
a constant and a tolerance value of 0.001 for optimization. For the {\em
  Ridge} algorithm, we used the same tolerance value and an alpha value of 1 as
regulation strength. For the SGD algorithm, we used the following settings: squared
loss, penalty L2, alpha value of 0.0001, L1 ratio of 0.15, epsilon as
0.1 and eta as 0.01. For the SVR model, we used RBF (Gaussian) kernel
with a $C$ value of 1000. For all the aforementioned algorithms, we
set the maximum number of iterations to 10,000. These models are
denoted as {\em 2-D poly: ElasticNet}, {\em 2-D poly: Lasso}, {\em 2-D
  poly: Ridge}, {\em SGD} and {\em SVR} in the result figures,
respectively.

The NN-based models are implemented using TensorFlow version
r1.12~\cite{tensorflow2015}. We considered both a fixed NN
structure and the automatically optimized NN structures for each
benchmark application as described in Section~\ref{sec:nn_models}, to
demonstrate the importance of hyperparameter optimization for NN
models. Here, the fixed structure had 5 fully-connected layers and 35
neurons per layer, which are the same the largest NN structure of the
hyperparameter optimization search space as defined in
Section~\ref{sec:nn_models}. This largest NN structure is chosen with
the assumption that a more complex NN model would be expected to provide
better prediction accuracy.


For hyperparameter optimizaton, we employed two libraries, HyperOpt
version 0.1.1~\cite{hyperopt} (for TPE optimization) and
Scikit-optimize version 0.5.2~\cite{scikitoptimize} (for Bayesian
Optimization). Both of the hyperparameter optimization libraries have
been set to 200 iterations for finding the high-accuracy parameters. Our
evaluations show that, increasing the number of iterations up to
1,000 will not significantly improve the prediction accuracy (less
than 2 percent) for the resulting NN models. However, with the
optimization time has a linear relation with the number of iterations,
the training time can increase by up to 5 times for 1,000
iterations. The resulting NN models are denoted as {\em NN:HyperOpt}
and {\em NN: SkOpt}, respectively.

\subsection{Validation of {\em uPredict}}
\label{sub:validation}

Based on the executions of two PARSEC benchmarks ({\em
  Streamcluster} and {\em Cannel}, which have higher prediction errors as shown later) on our private cloud, we first
validated the effectiveness of {\em uPredict}. Here, Figure~\ref{res:validation} shows the measured
(actual) execution times (the blue star points in the top figures) for
the two benchmarks as well as the corresponding number of background
VMs and their applications (in the bottom figures) for the duration of
24 hours. Clearly, the execution times of the benchmarks can vary
drastically (more than 10 times) due to variations in the severity of resource
contention caused by the background VMs and their applications on the
same host machine. Therefore, it is imperative to develop user-level
framework and tools for ordinary cloud users to get reasonably
accurate performance prediction for their applications and to support
their cost effective planning and operations.

\begin{figure*}
  \begin{center}
    \includegraphics[width=0.99\textwidth]{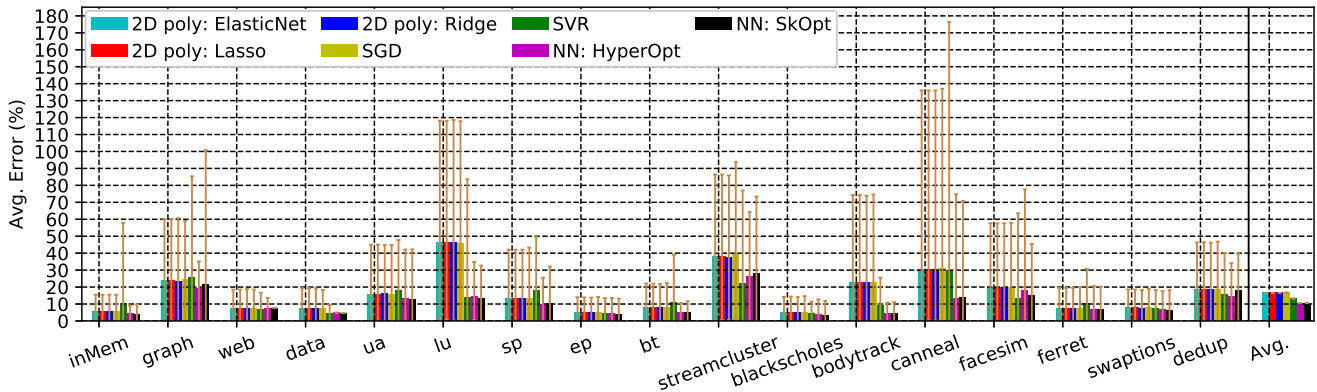}
  \end{center}
  \vspace{-.1in}
  \caption{Prediction errors of the predictive models in {\em uPredict} for the benchmark applications on the private cloud.}
  \label{res:errors-private}
\end{figure*}

\begin{table*}[ht]
\setlength{\tabcolsep}{2pt}
\vspace{-.1in}
\begin{center}
\caption{The standard deviation of the prediction errors (\%) for the benchmarks and predictive models on the private cloud.}
\label{tab:local-std}
\vspace{-.1in}
\begin{tabular}{|l|r|r|r|r|r|r|r|r|r|r|r|r|r|r|r|r|r|}

\hline
  & {\em inMem} & {\em graph} & {\em web} & {\em data} & {\em ua} & {\em lu} & {\em sp} & {\em ep} & {\em bt} & {\em sc} & {\em blackscholes} & {\em bodytrack} & {\em canneal} & {\em facesim} & {\em ferret} & {\em swaptions} & {\em dedup} \\ \hline
2D: ElasticNet     & 6.7 & 24.7 & 9.8 & 10.6 & 16.6 & 36.4 & 18.6 & 4.5 & 9.6 & 54.9 & 8.8 & 22.6 & 42.3 & 23.6 & 14.9 & 13.8 & 27.2 \\ \hline
2D: Lasso     &  6.7 & 24.7 & 9.9 & 10.6 & 16.6 & 36.4 & 18.6 & 4.5 & 9.6 & 54.9 & 8.8 & 22.6 & 42.3 & 23.6 & 14.9 & 13.6 & 27.2 \\ \hline
2D: Ridge     &  6.7 & 24.6 & 9.8 & 10.7 & 21.5 & 35.5 & 18.6 & 4.5 & 9.7 & 54.8 & 8.8 & 22.5 & 42.3 & 23.6 & 14.9 & 12.0 & 27.2 \\ \hline
2D: SGD       &  6.6 & 25.2 & 9.8 & 10.6 & 16.8 & 36.1 & 18.7 & 4.5 & 9.6 & 55.1 & 8.7 & 22.8 & 42.4 & 23.6 & 16.6 & 13.7 & 27.1 \\ \hline
SVR       & 17.9 & 29.2 & 9.6 & 10.6 & 18.4 & 27.8 & 14.9 & 4.6 & 12.8 & 28.0 & 12.5 & 10.2 & 62.1 & 28.6 & 24.0 & 14.7 & 26.3 \\ \hline
NN: HyperOpt  &  6.0 & 17.9 & 10.3 & 10.1 & 15.7 & 17.7 & 16.0 & 4.5 & 9.2 & 49.8 & 8.7 & 7.0 & 23.6 & 23.5 & 11.8 & 9.1 & 25.6 \\ \hline
NN: SkOpt     &  5.9 & 47.5 & 10.3 & 10.0 & 15.5 & 17.7 & 17.0 & 4.5 & 9.2 & 60.4 & 8.9 & 7.0 & 18.2 & 21.6 & 11.5 & 9.0 &  28.1 \\ \hline

\end{tabular}
\end{center}
\end{table*}

We can see from the figures that the execution times for {\em
  Streamcluster} and {\em Cannel} can be as low as around 90 and 30
seconds, respectively, at the beginning of each 2-hour interval. This
is due to the fact that, when the number of background VMs changes at
each 2-hour interval, the executions of the interfering applications
in all background VMs stop for the first 5 minutes, during which the
level of resource contention is rather low.

The predicted execution times (the red x points) utilizing the derived
SVR models for the two benchmarks are also shown in the top two
figures, respectively. Here, it is hard (if not impossible) to
associate a predicted execution time of the benchmarks with its
corresponding measured one in the figures. Although the predicted
execution times have several outliers for both benchmarks due to the
limitations of the predictive model, especially during the transition
period of changing background VMs and applications, it can be clearly
seen that the pattern (or trend) of the predicted execution times
closely follows that of the measured ones. Such patterns match the
severity of resource contention due to the background VMs and their
applications as shown in the bottom figures. Therefore, we can also
say that this experiment validates our hypothesis on the devised
micro-benchmarks, which can properly assess the resource contention in
the target VM at runtime.

\vspace{-.1in}
\subsection{Evaluation of {\em uPredict} in a Private Cloud}
\label{sub:private}

For the private cloud where the resource contention is
rather high with the controlled background VMs and applications,
Figure~\ref{res:errors-private} shows the the prediction errors of the
considered seven predictive models in {\em uPredict} for all the 17
benchmark applications. Here, the solid bars indicate the average and
the associated vertical lines show the 95-percentile of the prediction
errors. As explained earlier, for each benchmark, 80\% of the data
points were utilized to train/derive its application/VM-specific
predictive models while the remaining 20\% were used to test their
prediction accuracies.

First, for average prediction errors, the two NN-based predictive
models with hyper-parameters optimizations (i.e., {\em NN:HyperOpt}
and {\em NN:SkOpt}) perform the best for almost all benchmark
applications with lower than 20\% errors (except {\em streamcluster}
has 28\% error). Moreover, the overall average prediction errors by
considering all 17 benchmark applications were only 9.8\% for the
NN-based models. This indicates that the proposed uPredict with
NN-based models can indeed provide quite accurate performance
predictions for applications in a multi-tenant cloud environment even
with high resource contention (where our private cloud has up to 7
background VMs). However,\eat{we would like to point out that,} without
hyperparameter optimizations, the fixed structure (i.e., 5
layers and 35 neurons per layer) NN models can perform
rather worse with the overall average prediction errors being 60\% (and
up to 154\% prediction errors for some benchmark applications), which
is not shown in the figure.

\begin{figure*}
\begin{center}
 $\begin{array}{cc}
 \epsfxsize = 3in
  {\includegraphics[width=0.49\textwidth]{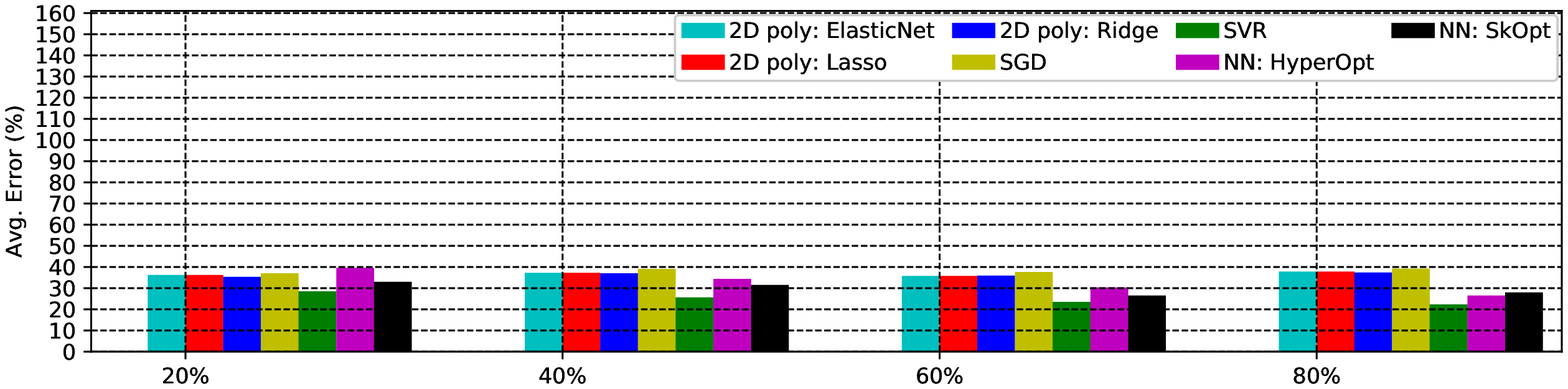}}
&\epsfxsize = 3in
  {\includegraphics[width=0.49\textwidth]{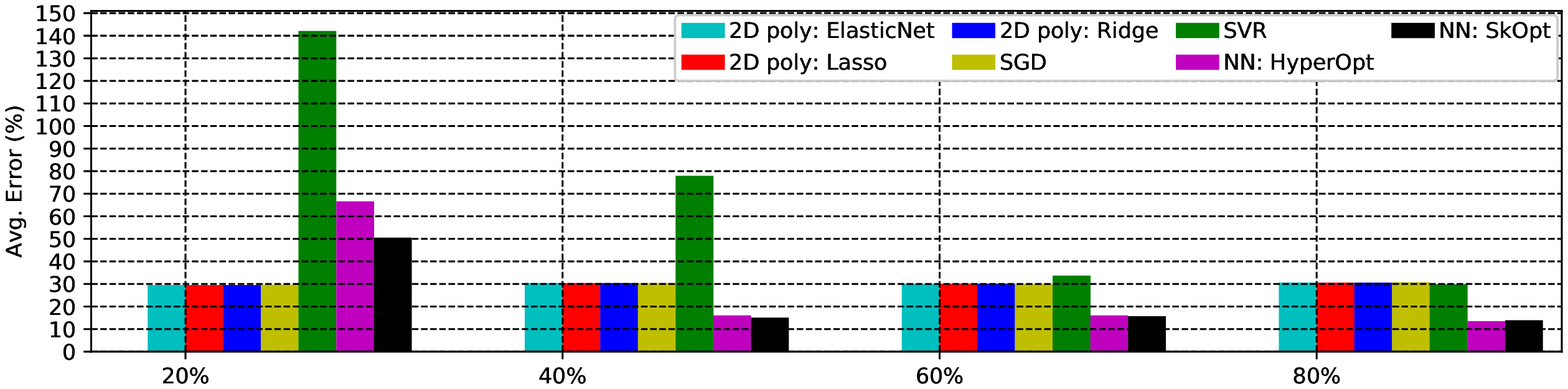}}\\
  \mbox{{\small a. Streamcluster }} 
& \mbox{{\small b. Canneal }} \\
\end{array}$
\end{center}
\vspace{-.15in}
\caption{Prediction errors of the predictive models with different training data for {\em Streamcluster} and
  {\em Canneal} on the private cloud.}
\vspace{-.1in}
\label{res:errors-vs-data}
\end{figure*}

For comparison, all four polynomial-regression based predictive models
perform relatively worse with the overall prediction errors for all
benchmarks being around 17\%, which is about 8\% higher that those of
the NN-based models. One possible reason for the worse performance of
the polynomial regression models compared to that of the NN-based
models is that, for certain applications, the relationships between
the profiling results from the micro-benchmarks and the actual
execution times of the applications are not necessarily
polynomial. Neural networks, on the other hand, have shown great
potential in finding relationships that are neither linear nor
polynomial~\cite{NeuralNonlinear}. For SVR models, while they
performed relatively better than the polynomial regression models for
most benchmark applications, their overall prediction accuracy is
still behind the NN-based models. The results indicated that SVR
models might be able to find non-polynomial relationships, however,
they are not as powerful as the NN-based models for predicting the
performance of cloud applications in clouds with high resource contention.

On the other hand, the prediction errors of the predictive models are
application dependent and vary in quite a large range. For several
applications (such as {\em web, data, ep, blackscholes} and {\em
  swaptions}), their 95-percentile prediction errors can be lower than
20\% for all the considered predictive models. However, for other
memory-intensive applications (such as {\em lu, streamcluster} and
{\em canneal}), their 95-percentile prediction errors can be more than
90\% for the polynomial regression predictive models. In particular,
for {\em canneal}, its 95-percentile prediction error for the SVR
model can be as high as 178\%. Such high prediction errors for the
outliers may have several causes. First, it can come from the
limitations of the predictive models and the micro-benchmarks that can
only profile a subset of factors that affect the performance of cloud
applications. We suspect that the relatively-high errors were caused
by the mismatch between the memory access patterns of the
micro-benchmarks and aforementioned benchmarks. They may also be
caused by the fact that the memory micro-benchmark is mainly profiling
off-chip memory contention, which does not include cache access
patterns.

Moreover, detailed analysis into the results shows that most of the
large errors were from predicted points when there were changes in the
background VMs and applications. During these changes, the
micro-benchmark can profile the resource contentiousness of the
startup or shutdown of background VMs. The benchmarks, however, were
later executed along with iBench applications, which have different
resource contentiousness. Consequently, the predicted results using
the profiled resource contention during VMs startups and shutdowns
were relatively inaccurate. Third, benchmark applications can have
different sensitivities to the contention of various underlying
hardware resources, which make some be more difficult to predict their
performance accurately than the others.

Table~\ref{tab:local-std} further shows the standard deviation of the
prediction errors for the predictive models and benchmark
applications. In general, the standard deviations for all the cases
are comparable to the average prediction errors, especially for the
applications that have larger error ranges (i.e., higher
95-percentile), such as {\em lu, streamcluster (sc)} and {\em
  canneal}. Such standard deviations indicate that it is difficult to
obtain accurate and stable predicted execution times for the
applications running in the cloud environment with high resource
contention. However, we would like to point out that, detailed analysis shows that 75\% of the predicted results have
errors no more than 10\% higher than the average prediction errors.

\vspace{.1in}
\noindent{\bf Effects of Training Data:} For the two representative
benchmarks, {\em streamcluster} and {\em canneal},
Figure~\ref{res:errors-vs-data} show their average prediction errors
for the predictive models when different amount of training data (up
to 80\%) is utilized. It can be see that, the four polynomial models
can achieve almost the same prediction performance for both
applications with only 20\% of training data being utilized. On the
other hand, for the SVR and NN models, having more training data can
generally improve their accuracies with reduced prediction errors
(especially for the case of SVR and {\em canneal}). The results for other
applications are similar, which are not shown due to space
limitation. Therefore, when the training data is limited, it may be
more beneficial to exploit the polynomial models instead of the
complex NN models.

\vspace{-.1in}
\subsection{Evaluation of {\em uPredict} in Public Clouds}
\label{sub:public}

\begin{figure*}
  \begin{center}
    \subfloat[Amazon EC2 Cloud.\label{res:errors-aws}]
             {\includegraphics[width=0.99\textwidth]{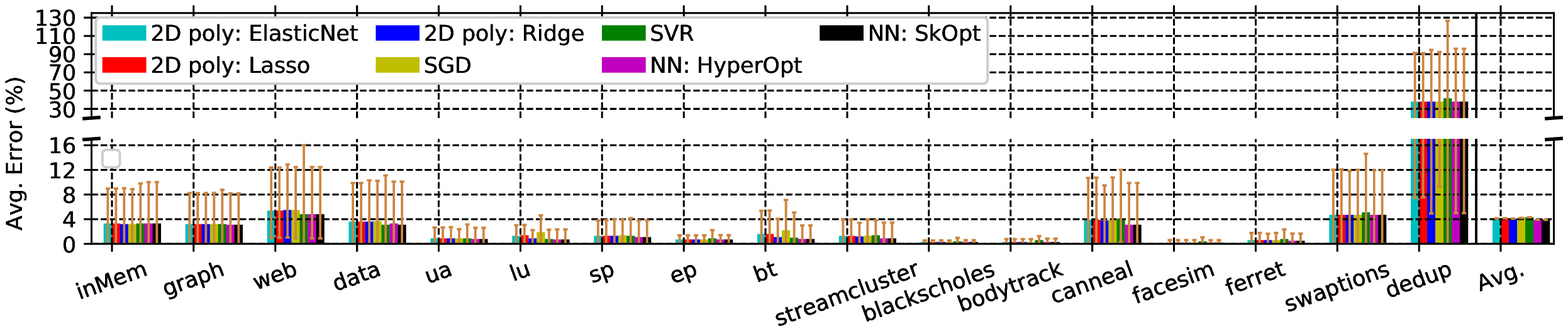}} \\
             \subfloat[Google Cloud Engine.\label{res:errors-gce}]
    {\includegraphics[width=0.99\textwidth]{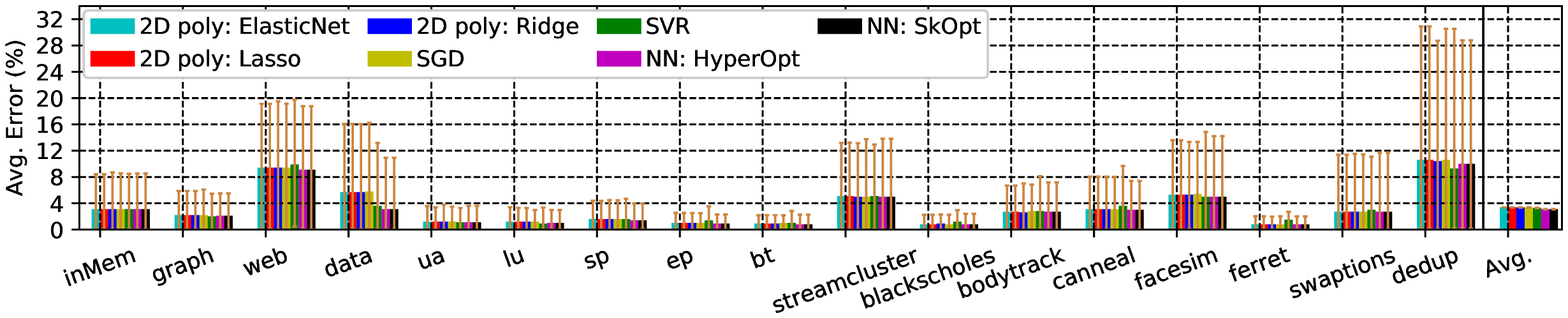}}             
  \end{center}
  \vspace{-.1in}
\caption{Prediction errors of the predictive models in {\em uPredict} for the benchmark applications on the public clouds.}
\label{res:errors-public}
\end{figure*}

Figure~\ref{res:errors-public} shows the average and 95-percentile
prediction errors of the studied predictive models for the benchmarks
on the two public clouds, Amazon AWS and Google GCE,
respectively. Here, Figure~\ref{res:errors-aws} first gives the
results on AWS, where all predictive models have the average
prediction errors less than 7\% and 95-percentile less than 14\% for
all benchmarks except {\em dedup}. Moreover, except for {\em dedup},
the average prediction errors of all the predictive models for all the
benchmarks differ by at most 3\%, which suggests that the predictive
models have almost equivalent accuracy on AWS. The lower prediction
errors in AWS than the private cloud are mainly due to the relatively
low resource contention. The execution times of the benchmarks on AWS
can fluctuate up to 25\%, compared to up to 10 times
fluctuation in the private cloud.

These results show that the proposed profiler-based framework with any
considered predictive model can predict most single-VM benchmarks
quite accurately on AWS. Hence, {\em uPredict} is feasible for
ordinary AWS users to obtain accurate performance prediction without
the need of the exact knowledge or control of the underlying execution
environments. Although the NN-based models can provide slightly higher
prediction accuracy, it takes much more time to train compared to that
of polynomial regression based models, especially with the
hyperparameter optimizations.

\eat{Moreover, as the figure shows regression techniques and
SVR are highly comparable to Neural Network algorithm in terms of
accuracy in public cloud infrastructure. However, results show that
the NN models will provide a higher accuracy. But higher accuracy
results requires more computation power. It take about 5 second to
train the optimal model and about 0.005 second for predicting the
results, which is about 400\% and 500\% more in training and
prediction time compare to regression algorithms.
}

The large prediction error with {\em dedup} was caused by OS thread
scheduling issue rather than the models mispredicting the impact of
resource contention. We observed that {\em dedup} has two execution
times on AWS, which are either about 5 seconds or about
11 seconds. Such an execution pattern is usually caused by issues
other than contention (if it is caused by contention, then the
execution times would be spread between these two times). Further
analysis revealed that {\em dedup} created about 30 to 40 threads even
when specified to use only 16 worker threads. These many threads
prevented the OS from providing a stable scheduling behavior on the
16-core VM, resulting in two groups of execution times. When we
reduced the worker thread count to below $8$ (which in turn reduced
the total thread count), the thread scheduling was more stable, and
the proposed methodology could achieve an average error less than 10\%
for all models. However, for consistency, the 16-worker-thread results
were used in the figures.

Figure~\ref{res:errors-gce} further shows the results on GCE. Here,
the average and 95-percentile prediction errors from all predictive
models for all benchmark applications are no more than 10\% and
38\%. In fact, except for {\em dedup}, the 95-percentile prediction
errors for all other benchmarks are less than 20\%. The overall
average prediction errors for all the benchmarks are less than 4\% for
all the studied predictive models. These results show that the
proposed predictive framework is highly accurate on GCE for the
considered benchmark applications as well. Combining with the findings
from AWS, these results further confirm that it is feasible for
ordinary cloud users to utilize uPredict to get accurate performance
prediction on public clouds without the need of the exact knowledge or
control of the underlying execution environments.

Moreover, similar to the results on AWS, the average prediction errors
from all predictive models for the benchmarks differ by at most 2\%,
suggesting that these predictive models have almost equivalent
performance in terms of prediction accuracy on GCE. Also, the same as
in AWS results, the figure shows that the polynomial regression and
SVR models are comparable to neural network models in terms of
prediction accuracy, although on average the NN models provide
slightly higher accuracy.

\begin{figure*}
  \begin{center}
    \subfloat[{\em StreamCluster}.\label{fig:sens_stream}]
              {\includegraphics[width=1\textwidth]{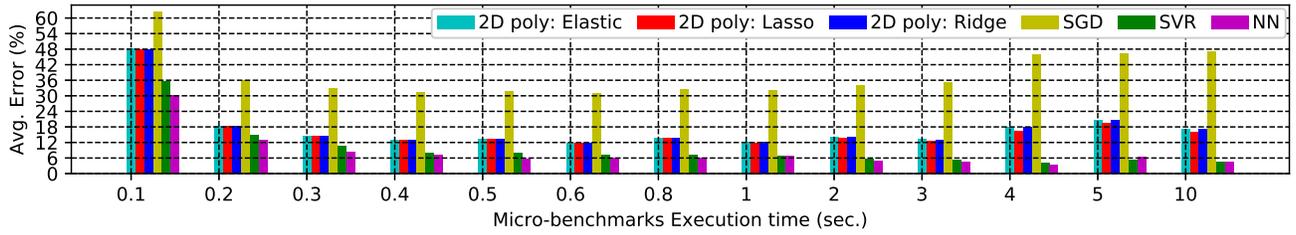}}\\
                  \subfloat[{\em Canneal}.\label{fig:sens_canneal}]
              {\includegraphics[width=1\textwidth]{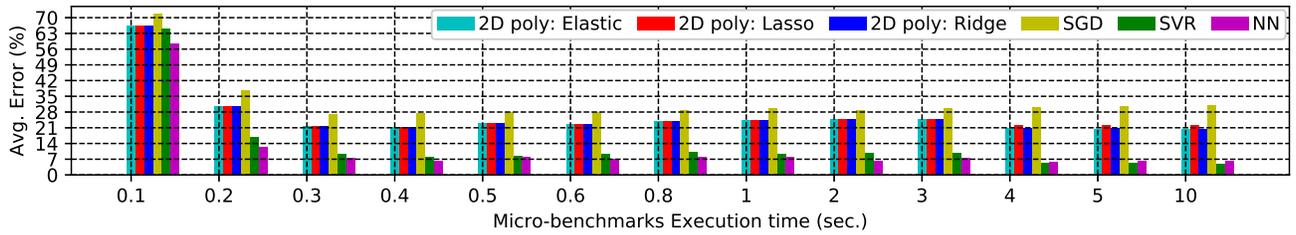}}\\
                  \subfloat[{\em Swaptions}.\label{fig:sens_swaptions}]
              {\includegraphics[width=1\textwidth]{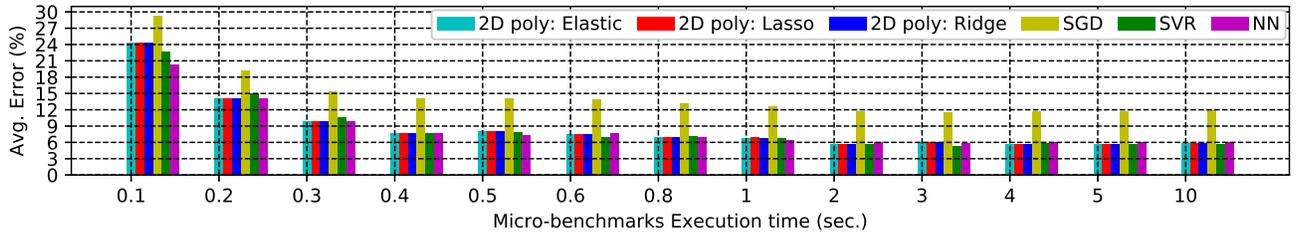}}\\
                 \subfloat[{\em ep}.\label{fig:sens_ep}]
              {\includegraphics[width=1\textwidth]{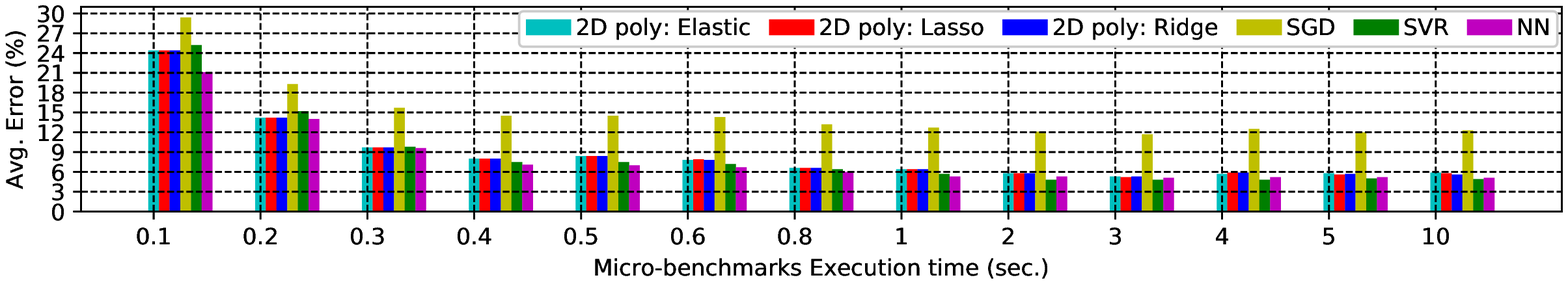}}\\
                  \subfloat[{\em graph}.\label{fig:sens_graph}]
              {\includegraphics[width=1\textwidth]{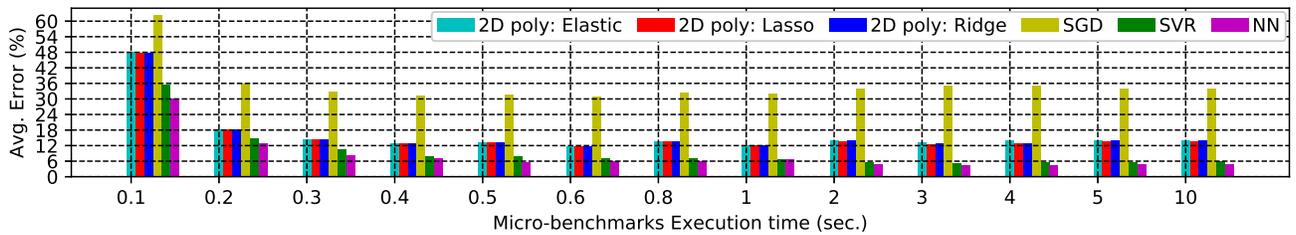}}\\
              
  \end{center}
  \vspace{-.1in}
\caption{Prediction accuracy sensitivity to profiling duration.}
\label{fig:microbench-sensivity}
\end{figure*}

\subsection{Sensitivity to Profiling Execution Length}\label{sec:profiling_sensitivity}
As stated in Section~\ref{sec:duration}, the lengths of the profiling executions are important parameters for uPredict, as uPredict relies on the micro-benchmarks to provide accurate measurement of current severity of contention. Although we have been using 3-seconds of profiling executions, a smaller profiling length may already provide good knowledge on the level of contention. To evaluate the impact of profiling length on prediction accuracy, we conducted an sensitivity experiment using 5 representative applications on our private cloud. 

In this sensitivity experiment, micro-benchmarks were executed with different execution time of 0.1 sec to 10 seconds before the execution of target application. Each of the five applications were executed for 800 iterations along with the micro-benchmarks. Except for the profiling length, the experiment setup was the same as discussed in Section~\ref{sub:private}. As described in the uPredict methodology, the application's execution times and micor-benchmark performance were collected during this experiement. After the execution, 80\% of the 800 iterations' data (i.e., 640 data points) were used to build the uPredict prediction models using the performance results of the micro-benchmarks with different profiling lengths. These models were then tested on the rest 20\% (i.e., 160) data points.

Fig.~\ref{fig:microbench-sensivity} shows the accuracy of these prediction models for all 5 representative benchmarks using different prediction algorithms and profiling execution lengths. As Fig.~\ref{fig:microbench-sensivity} shows, after profiling length of 0.4 seconds, the accuracy improvement from longer profiling runs were limited, whereas after three seconds, there was nearly accuracy improvements. These results suggested that a profiling length of 0.4 seconds may be long enough to obtain good accuracy, while there is generally no need to profiling for more than 3 seconds.



\subsection{Case Study: Load-Balancing with {\em uPredict}}
\label{sub:casestudy}

To illustrate the usage of {\em uPredict}, we have conducted a case
study of load-balancing for two cloud servers. Here, each cloud server
is a VM with the same configuration as the one in previous experiments
(i.e., 16 VCPUs and 64 GB memory) and both run under OpenStack on two
separate host machines. Each machine has two Intel Xeon E5-2630
processors (for a total of 16 cores) and 128GB memory. In addition to
the VM acting as one cloud server, up to three (3) background VMs may
be created randomly to run applications from iBench on each host
machines. These background VMs/applications change at different fixed
intervals on the machines.

We assume that a user-level load-balancer is adopted to direct the
requests of cloud users to run the benchmark {\em Graphics Analytic}
of CloudSuite to one of the two cloud servers at runtime. Three
different load-balancing schemes were investigated in this
study. First, the {\em dummy} load-balancer just distributes the
received user requests {\em alternatively} to the two cloud servers
one after the other without any information from the cloud servers
being considered. Second, a simple {\em queue-based} load-balancer
considers the number of requests in the waiting queues of both cloud
servers and distributes a new user request to the server with a
shorter waiting queue. Finally, the {\em uPredict-based} smart
load-balancer considers the predicted execution times for the requests
in the waiting queues based on the current profiled resource
contention from the micro-benchmarks on both cloud servers. A new user
request would be distributed to the server where the request is
expected to complete earlier.

The first experiment was conducted for 3 days, where the user requests
were periodically sent to the load-balancer starting with 12 requests
per hour. The rate of incoming user requests gradually increases to 24
requests per hour and then decreases to 12 requests per hour in the
end, where the request rate changes after every 4 hours for a duration
of 3 days (i.e., 72 hours). With the resource contention from the
background VMs/applications, the execution times for {\em Graphics
  Analytic} have the range from 45 seconds to about 7 mintues. The
above request rate is rather high (denoted as {\em high-load})
especially at the peak rate of 24 requests per hour.

\begin{table}[tbp]
\caption{The average turnaround and execution times (seconds) of {\em Graphic Analytic} under different balancers for {\em high-load}.}
\label{tab:case-study-high}
\vspace{-.1in}
\begin{center}
\begin{tabular}{|l|c|c|} \hline
 balancers & turnaround time (s) & execution time (s) \\ \hline\hline
 dummy-alternate  & 3563 & 312 \\ \hline
 
 queue-based      &  987 & 276 \\ \hline
 
 uPredict-based   & 1066 & 256 \\ \hline

\end{tabular}
\end{center}
\end{table}

Table~\ref{tab:case-study-high} shows the average turnaround and
execution times for the generated requests of running {\em Graphic
  Analytic} under different load-balancers. Clearly, without
considering resource contention and workload (i.e., queue length)
on the cloud servers, the dummy-alternate scheme can result in very
high turnarond time, which is more than 3 times of those for the
queue-based and uPredict-based schemes. Although the queue-based
scheme does not consider the resource contention,
the queue length actually implies the delivered performance on each
server. Hence, the average turnaround times for the queue-based and
uPredict-based schemes are quite close.

For the average execution time of the requests, by avoiding the cloud
server with high resource contention (implied by its queue length),
the queue-based scheme can improve it for about 11.5\% over
dummy-alternate. Given that the uPredict-based scheme tries to
execute most requests on the server that can deliver higher
performance with less resource contention, its average execution
time can be improved by 18\% compared to that of dummy-alternate.

In the second 3-day experiment, we reduced the request rate by half
through the duration (denoted as {\em low-load}), and the results are
shown in Table~\ref{tab:case-study-low}. In this case, as the waiting
queues on both cloud servers are empty for most of the time, the
queue-based scheme performs relatively worse, where both of its
average turnaround and execution times of the requests are around 12\%
better than those of the dummy-alternate scheme. By exploiting the
resource contention on the cloud servers, the uPredict-based scheme
can further improve 19\% and 10\% over the queue-based scheme for the
average execution and turnaround times of the requests,
respectively. Note that, the profiling overheads of running the
micro-benchmarks in the uPredict-based scheme were already included in
the resulting turnaround times of the generated requests.

\begin{table}[htp]
\caption{The average turnaround and execution times (seconds) of {\em Graphic Analytic} under different balancers for {\em low-load}.}
\label{tab:case-study-low}
\vspace{-.1in}
\begin{center}
\begin{tabular}{|l|c|c|} \hline
 balancers & turnaround time (s) & execution time (s) \\ \hline\hline
 dummy-alternate  &  360 & 339 \\ \hline
 
 queue-based      &  318 & 300 \\ \hline
 
 uPredict-based   &  287 & 242 \\ \hline

\end{tabular}
\end{center}
\end{table}

\eat{
  
\begin{table}[htbp]
\caption{Average Algorithms overhead}
\label{table:overhead}
\vspace{-.1in}
\begin{center}
\begin{tabular}{|l|c|c|} \hline
 Algorithm & Training Overhead & Prediction overhead \\ \hline\hline
 2D poly: ElasticNet   & 0.180s & 0.0001s\\ \hline
 
 2D poly: Lasso    & 0.182s &  0.0001s \\ \hline
 
 2D poly: Ridge  & 0.184s & 0.0001s \\ \hline
 
 SGD   & 0.740s & 0.0001s \\  \hline
 
 SVR   & 0.033s & 0.0040s \\ \hline
 
 NN    & 5.6s & 0.0042s \\  \hline 
 
\end{tabular}
\end{center}
\end{table}

}
 